\documentclass[trackchanges]{aastex631}

\shorttitle{MOS of Galaxy Clusters at $z \sim 0.95$ in UDS}
\shortauthors{Lee et al.}
\graphicspath{{./}{figures/}}

\usepackage{graphicx}

\begin{document}

\title{Multi-Object Spectroscopy of Galaxy Clusters at $z \sim 0.95$ in Ultra Deep Survey Field with Different Star-formation Properties and
Large-scale Environments}

\correspondingauthor{Seong-Kook Lee}
\email{s.joshualee@gmail.com}

\correspondingauthor{Myungshn Im}
\email{myungshin.im@gmail.com}

\author{Seong-Kook Lee}
\affiliation{SNU Astronomy Research Center, Department of Physics and Astronomy, Seoul National University, Seoul, Korea}
\affiliation{Astronomy Program, Department of Physics and Astronomy, Seoul National University, Seoul, Korea}

\author{Myungshin Im}
\affiliation{SNU Astronomy Research Center, Department of Physics and Astronomy, Seoul National University, Seoul, Korea}
\affiliation{Astronomy Program, Department of Physics and Astronomy, Seoul National University, Seoul, Korea}

\author{Bomi Park}
\affiliation{SNU Astronomy Research Center, Department of Physics and Astronomy, Seoul National University, Seoul, Korea}
\affiliation{Astronomy Program, Department of Physics and Astronomy, Seoul National University, Seoul, Korea}

\author{Minhee Hyun}
\affiliation{Korea Astronomy and Space Science Institute, Daejeon, Korea}

\author{Insu Paek}
\affiliation{SNU Astronomy Research Center, Department of Physics and Astronomy, Seoul National University, Seoul, Korea}
\affiliation{Astronomy Program, Department of Physics and Astronomy, Seoul National University, Seoul, Korea}

\author{Dohyeong Kim}
\affiliation{Department of Earth Sciences, Pusan Natioanal University, Busan, Korea}


 
\begin{abstract}

While galaxy clusters are dominated by quiescent galaxies at local, they show a wide 
range in quiescent galaxy fraction (QF) at higher redshifts. Here, we present 
the discovery of two galaxy clusters at $z \sim 0.95$ with contrasting QFs despite 
having similar masses (log ($M_{200}$/$M_{\odot}$) $\sim 14$) and spectra 
and redshifts of 29 galaxies in these clusters and 76 galaxies 
in the surrounding area. 
The clusters are found in the Ultra Deep Survey (UDS) field and confirmed through multi-object 
spectroscopic (MOS) observation using the Inamori Magellan Areal Camera and Spectrograph 
(IMACS) on the Magellan telescope. The two clusters exhibit QFs 
of $0.094_{-0.032}^{+0.11}$  and $0.38_{-0.11}^{+0.14}$, respectively. 
Analysis of large-scale structures (LSSs) surrounding these clusters finds 
that properties of these clusters are consistent with the anti-correlation trend between 
the QF and the extent of surrounding LSS, found in \citet{lee19}, which 
can be interpreted as a result from the replenishment of young, star-forming galaxies keeps the QF low when galaxy clusters 
are accompanied by rich surrounding environments.

\end{abstract}



\section{Introduction} \label{sec:intro}

Understanding the evolution of star formation properties of galaxies has profound implications 
for the study of galaxy evolution and cosmology.  
A galaxy's star formation history (SFH) is intricately tied to its overall evolution, 
with different galaxy types exhibiting distinct patterns in their SFHs \citep[e.g.][]{lon99,cle06,fit15}.
Tracing the evolution of star formation properties provides valuable insights into the processes 
governing the transformation and evolution of galaxies from one type to another \citep{ken98,hea04,ren09}.
Also, the star formation rate (SFR) in galaxies has varied across cosmic time. 
Examining this variation allows us to deduce the cosmic star formation history (SFH) 
of the universe, unveiling when the peak of star formation occurred and how it has evolved over time 
\citep{mad96,har04,hea04,mad14}. 
The evolution of star formation properties also offers valuable constraints for cosmological models. 
For instance, it can aid in testing predictions related to dark matter, dark energy, and the large-scale 
structure (LSS) of the universe. 
Comparing observed SFRs with theoretical predictions from models proves beneficial in refining our 
comprehension of cosmological models \citep[e.g.][]{hen15,kne15,som15,vog20}.

In this regards, it is particularly crucial to investigate how star formation in galaxies evolves or ceases. 
Understanding the processes governing star formation evolution unravels the lifecycles of galaxies. 
Galaxies undergo various development stages, transitioning from active star-forming phases to quiescent states. 
Studying these transitions provides insights into the factors driving galaxies to change over cosmic time.  

Star formation is a intricate process that entails interactions among gas, dust, and stars. 
Feedback mechanisms, such as stellar winds and supernovae, play a crucial role 
in regulating star formation by expelling gas and heating the interstellar medium. 
Consequently, understanding both the cessation of star formation and the intricate connection 
between star formation and feedback processes is essential for gaining insights into 
how galaxies autonomously regulate their growth and how these processes influence various galaxy properties.

The proposed quenching mechanisms can generally categorized into two types: internal and external. 
The internal mechanisms are reflected in  the correlation between the star formation rate (SFR) and 
stellar mass \citep{kau03,jim05,bal06}, while external mechanisms are manifested in the correlation 
between the SFR and the environment \citep{lew02,kau04,bal06}. 
Numerous studies have delved into the relative importance and effects of these internal and external quenching 
processes \citep[e.g.,][]{pen10,lee15,cha20,gu21}.

In the present-day universe, many galaxies in dense environments exhibit little to no ongoing star 
formation activity, leading to a distinct environment-dependent bimodality.
However, observational evidence indicates that this environment-dependent bimodality weakens 
at higher redshifts. 
By redshift $z \sim 1$, this environmental dependence of galaxy properties becomes notably weak \citep{sco13}.
Additionally, the gap in quiescent galaxy fraction (QF) between clusters and the field also 
diminishes with increasing redshift \citep[e.g.,][]{lee15,pap18,lem19,sar21}.
This suggests that galaxies in clusters are actively forming stars, similar to their counterparts 
in the field at this redshift, in contrast to the local universe.

To understand the diminishing environmental dependence with increasing redshift, 
it is essential to explore the evolution of star formation properties within the context 
of galaxy cluster formation and evolution.
Galaxies that fall into a cluster undergo various physical quenching mechanisms, 
such as ram pressure stripping \citep{gun72, aba99}, strangulation \citep{lar80, bal00, pen15} 
and galaxy harassment \citep{moo96}, operating on different timescales. 
It is plausible that galaxies recently entering clusters at high redshift \edit1{($z > 1.5$)} may 
not have had sufficient time to undergo quenching.

Moreover, alongside the reduced environmental dependence on star formation activity 
in galaxies at $z \sim 1$, individual galaxy clusters exhibit significant variation in their star formation 
status, such as QF \citep{lee15,alb16,hay19,lee19}. 
However, it remains unclear what factors contribute to the observed cluster-by-cluster variation 
in the star formation properties of high-redshift ($z > 0.5$) galaxy clusters.

At high redshift, many galaxy clusters are still growing by accreting gas and galaxies from 
their surrounding environment. 
These accreted materials may contribute to maintaining a large fraction of star-forming galaxies 
within the clusters \citep{ell01,tra05,lub09,lee19}. 
In our recent study  \citep{lee19} conducted in the Ultra Deep Survey \citep[UDS;][]{alm07} field, 
we demonstrated a correlation 
between the QF of a galaxy cluster and the extent of the large-scale structures connected to the cluster.
This suggests that the large-scale environment surrounding galaxy clusters, acting as a reservoir of gas and 
star-forming galaxies, can influence the star-formation status of clusters.
In \citet{lee19}, we proposed the web-feeding model (WFM) to explain this phenomenon.

The aforementioned discoveries underscore the necessity for a thorough investigation into 
the star formation properties and the surrounding environment of high-redshift galaxy clusters. 
The research presented in this paper reflects our continuous efforts to understand the star 
formation status of galaxy clusters and their associated large-scale structures.

In this study, we present two newly spectroscopically confirmed galaxy 
clusters at $z \sim 0.95$ in the UDS field and provide redshifts for 105 galaxies within these clusters 
and the surrounding area.
Our aim is to investigate the star formation properties and characteristics of the 
surrounding large-scale structures associated with these clusters. 
This redshift range has been recognized as a critical period for the environmental quenching of 
star formation activity in galaxies \citep[e.g.,][]{pen10,sob11,qua12,lee15,lu21,mcn21}.

In Section \ref{sec:dat}, we offer a detailed explanation of the data, including cluster sample, 
spectroscopic observations, data reduction procedure, and redshift measurements. 
Utilziing the spectroscopic data and redshift measurements, we confirm the clusters and explore 
the properties of both the clusters themselves and their member galaxies in Section \ref{sec:res}.
Section \ref{sec:anal} is dedicated to presenting the correlation between the QF and the extent of 
the connected large-scale structure surrounding these clusters. 
Additionally, we discuss the QFs of these clusters, considering their dependence on stellar mass.

We adopt a standard cosmology with ($\Omega_{m}$, $\Omega_{\Lambda}$) = (0.3, 0.7) and 
$H_{0}$ = 70 km s$^{-1}$ Mpc$^{-1}$, consistent with observations over the past decades 
\citep[e.g.][]{im97}.
All magnitudes are given in the AB magnitude system \citep{oke74}. 

\section{Data and Sample}
\label{sec:dat}

\subsection{Photometric Data}
\label{sec:phodata}

\subsubsection{Multiwavelength Data}
\label{sec:multi}

The UDS field carries a wealth of multi-band photometric data collected from various facilities 
across approximately 0.8 deg$^{2}$. 
In this study, we utilize optical data from the Subaru telescope, near-infrared (NIR) data 
from the United Kingdom InfraRed Telescope (UKIRT), and mid-infrared (MIR) data 
from the Infrared Array Camera (IRAC) on the Spitzer telescope for 
conducting photometric redshift calculation and estimating stellar population properties 
through spectral energy distribution (SED) fitting.

The optical broad-band photometric data are obtained from the SUPRIMECAM on the Subaru 
telescope \citep[SXDS;][]{fur08}, with 3-$\sigma$ magnitude limits at each band as 
$B=28.4, V=27.8, R = 27.7, i' = 27.7$, and $z'=26.7$.
NIR photometric data are from the WFCAM on the UKIRT \citep[UKIDSS;][]{law07}, with 
5-$\sigma$ magnitude limits of $J=25.2, H=24.7$, and $Ks=24.9$.
The MIR data are from the Spitzer/IRAC as the SpUDS Spitzer Legacy Survey (PI: Dunlop), 
with depths of $\sim 24$ mag in channels 1 and 2.
For the Subaru/SUPRIMECAM and UKIRT/WFCAM data, auto magnitudes are used, 
while aperture-corrected total magnitudes are used for the Spitzer/IRAC data.
For further information on the photometry and the reliability of using the auto magnitudes, 
refer to \citet{lee15}.

In addition to these data sets, we use MIR data from the Multiband Imaging Photometer for Spitzer 
(MIPS) to explore dusty star-forming galaxies. 
As mentioned above, the UDS region was covered by the SpUDS survey, which includes 
MIR data from Spitzer/IRAC as well as MIPS 24 $\mu$m. 
The depth of MIPS 24 $\mu$m is $\sim 80 \nu$Jy, which corresponds to 
SFR $\sim 20$ M$_{\odot}$/yr at $z \sim 1$ \citep{cap06}.  
In this study, we investigate the MIPS image data and catalog within the Magellan-observed region 
to identify any MIPS sources (i.e. dusty SF galaxies) within and around the cluster region. 


\subsubsection{Photometric Redshift Estimation and SED Fitting}
\label{sec:sed}

Utilizing the aforementioned multiwavelength data, we estimate the photometric redshifts 
and stellar population properties, such as stellar mass, and SFR, for the galaxies 
in our sample, as explained in \citet{lee15} in detail. 
In brief, we utilized optical, NIR, and MIR broadband data to estimate the photometric 
redshifts using EAZY \citep{bra08}. 
We employ the $K$-band magnitude prior, as it yields slightly improved accuracy 
in photometric redshifts. 

To derive the stellar population properties of galaxies, we used our own 
SED-fitting code \citep{lee09} assuming delayed star-formation histories (SFHs) \citep{lee10}, 
with a functional form to describe the star formation rate at a given time $t$ as 

\begin{equation}
    \Psi (t, \tau) \propto \frac{t}{\tau^{2}} e^{-t/\tau},
\label{eq:sfheq}
\end{equation}

where $t$ represents the time since the onset of star formation and $\tau$ is the timescale 
parameter that governs the rate at which the SFR reaches its peak value.

The delayed SFH has been shown to well represent the actual SFHs of high-redshift galaxies 
\citep[e.g.][]{pac13,lee14}.

For the SED-fitting, we employ $\sim 4000$ spectral templates derived from the 
updated version of the BC03 model \citep{bru03}, assuming the \citet{cha03} initial mass function (IMF). 
We incorporate internal dust attenuation using the Calzetti law \citep{cal00}, 
while the intergalactic medium (IGM) dust attenuation follows the Madau law \citep{mad95}. 
After estimating the stellar population properties, we apply a stellar mass cut (log ($M_{*}$/M$_{\odot}) \ge$ 9.1) 
to our sample.  

Through the SED-fitting process, various parameters are derived for each galaxy, including 
stellar mass ($M_{*}$), age ($t$), star formation timescale ($\tau$), metallicity ($Z$), 
and color excess ($E(B-V)$).
For more detailed information regarding photometric redshift estimation and our SED-fitting process 
for our sample galaxies, please refer to \citet{lee15}. 


\subsection{Spectroscopy}
\label{sec:spec}

\subsubsection{Target Selection}
\label{sec:tar}

In \citet{lee15}, we identified galaxy cluster candidates within the redshift range of 
$0.5 \leq z \leq 2.0$ in the UDS field. 
These cluster candidates were identified based on photometric redshifts, 
identifying structures with galaxy number densities exceeding the $4 \sigma$ level. 
Please refer \citet{lee15} for more details about our cluster candidate selection method.  
The reliability of our cluster selection method has been well demonstrated in our previous work 
\citep{lee19} and also in \citet{sar21}.

From the galaxy cluster candidate list in \citet{lee15}, we selected several cluster 
candidates for follow-up spectroscopic observations, 
among which the observation and analysis of two $z \sim 0.95$ cluster candidates 
is presented in this paper. 
These two cluster candidates were chosen due to their distinct star formation properties, 
despite being at similar redshifts and separated by only $\sim 4$ Mpc. 

Subsequently, we identified galaxies within and surrounding these selected cluster candidates 
for spectroscopic follow-up observations.
The selection of target galaxies was based on their photometric redshifts, $R$-band magnitudes, 
and proximity to the candidate clusters. 
During the design of masks for the multi-object spectroscopic (MOS) observation, 
priority values were assigned to galaxies based on these specific galaxy properties.
Galaxies with brighter $R$-band magnitudes and closer proximity 
to the candidate clusters were given higher priority.


\subsubsection{Magellan Observation} 
\label{sec:obs} 

We conducted observations of two galaxy cluster candidates and their surrounding regions 
using the Inamori-Magellan Areal Camera and Spectrograph (IMACS) on the Magellan/Baade telescope. 
These two candidate clusters are at similar redshifts of $z \sim 0.95$ and are separated by $\sim 4$ Mpc.

employing the Magellan/IMACS f/2 camera, we performed multiobject spectroscopic (MOS) observations 
on twice occasions -- 
firstly on August 27, 2016 (2016B semester) and subsequently on August 31, 2019 (2019B semester). 
179 out of total 538 slits were assigned within $9' \times 18'$  sky region 
($34.0\,^{\circ} - 34.15\,^{\circ}$ in R.A., and -$5.0\,^{\circ} - $-$4.7\,^{\circ}$ in declination), 
covering the cluster candidates and their surrounding regions. 
The other 359 slits were allocated in the remaining mask area, targeting different cluster candidates 
and LSSs at different photometric redshifts. 
The spectroscopic data obtained from these remaining slits will be presented in our future work. 
For each mask, we assigned a slit for a standard ($K$-type) star for flux calibration.

The slit length and width were 6 arcsec $\times 1$ arcsec. 
The exposure times for each run were 2 hours for the 2016B run and 1.5 hours for the 2019B run.
In both runs, the exposure time was divided into four and three 30-min exposures, respectively. 
The average seeing were 1.5 arcsec for the 2016B run and 1 arcsec for the 2019B run.
We utilized a Grism with 200 lines per millimetre and the WB6300-9500 filter for the 2016B run, 
and the WBP5694-9819 filter for the 2019B run, 
providing wavelength coverage of 630$\sim$950 nm (570$\sim$980 nm) 
with a spectral resolution of R $\sim$ 650. 
These settings ensured adequate coverage of major spectroscopic features of 
the galaxies under investigation.

\begin{figure}
 \plotone{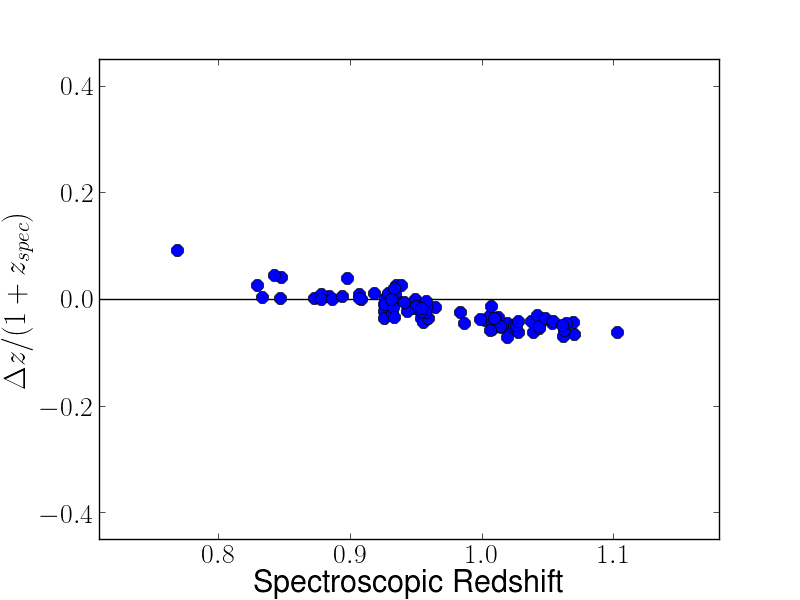}
 \caption{Photometric redshift error ($\Delta z / (1+z)$) as a function of spectroscopic redshift. 
Photometric redshift errors are small in the range of $0.8 < z < 1.0$.  \label{fig:szdz}}
\end{figure}

\begin{figure}
 \plotone{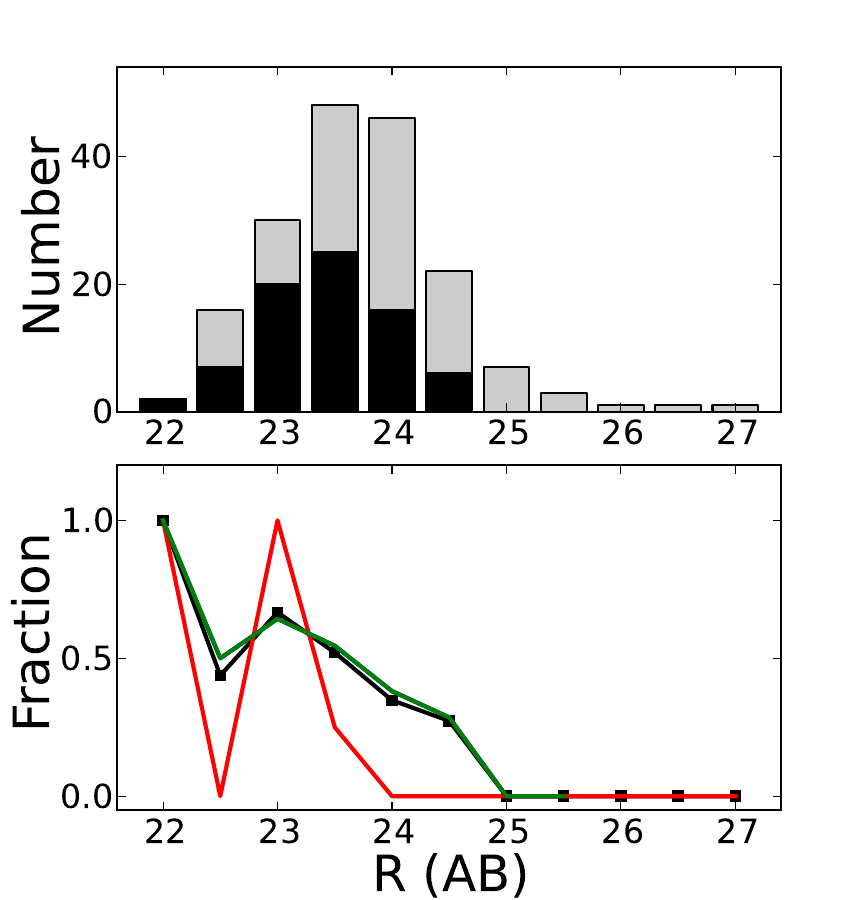}
 \caption{{\bf (Top)} Distributions of $R$ magnitude of our target galaxies (grey histogram) and of galaxies 
whose spectroscopic redshifts are measured with flag `a' (black histogram). 
{\bf (Bottom)} $R$ magnitude-dependent redshift measurement rate for blue ($R$--$z' < 1.5$; green line), 
red ($R$--$z' > 1.5$; red line) and total (black symbols and black line) galaxies with flag `a'.  \label{fig:specrat}}
\end{figure}


\subsubsection{Data Reduction}
\label{sec:redu}

Data reduction was performed using the latest version of the Carnegie Observatories
System for Multi-Object Spectroscopy software (COSMOS2), which 
is the software package provided by Carnegie Observatories for 
reduction of spectroscopic data obtained with the IMACS instrument.
We followed the same reduction procedure as outlined in \citet{lee19}.
For wavelength calibration, He-Ne-Ar arc frames were used.
Following the standard COSMOS2 routine, flat fielding and bias subtraction 
were conducted on the science frames. 
After sky subtraction, two-dimensional (2D) spectra were extracted and 
combined into a single 2D spectrum, 
The bulk of cosmic rays were removed in this process. 
For further details, please consult the COSMOS cookbook \citep{oem17}.
One-dimensional (1D) spectra were subsequently extracted from the 
combined 2D spectrum, employing an extraction width corresponding 
to twice the average seeing values. 
Flux calibration was done using a standard star --- the same star 
for both masks --- included in each mask.
Flux-calibrated spectra are available as machine-readable tables, 
with an example presented in Appendix A (Table~\ref{tab:calflx}).

\begin{figure}
 \plotone{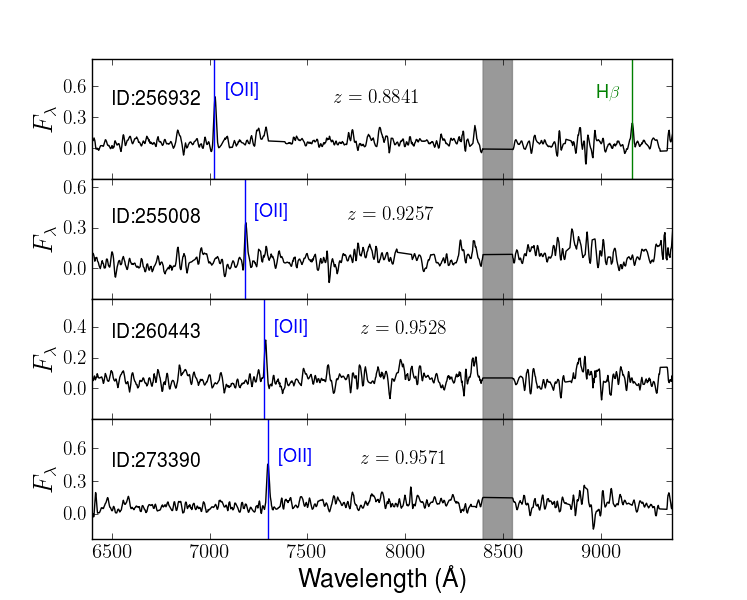}
    \caption{Example spectra of galaxies with redshift measurements obtained 
from our Magellan/IMACS observation. Among these galaxies, two are at $z \sim 0.95$ and 
are associated with UDS-OD5. 
The $y$-axis is in the unit of $10^{-17}$ $erg$ $s^{-1}$$cm^{-2}$\AA$^{-1}$. \label{fig:spexam}}
\end{figure}


\subsubsection{Redshift Measurement}
\label{sec:redsh}

From the extracted 1D spectra, galaxy redshifts were determined using the SpecPro software \citep{mas11}. 
Key spectral features utilized for redshift identification of galaxies within the redshift range of $0.8 < z < 1.1$ 
are [\ion{O}{2}] $\lambda$3727, H$\beta$, H$\gamma$ , and [\ion{O}{3}] emission lines,  
along with Ca H$\&$K and $G$-band absorption lines. 

For many galaxies at the redshift of cluster candidates ($z \sim 0.95$), redshift 
measurements rely on a single emission line feature, [\ion{O}{2}] $\lambda$3727. 
Despite this reliance on a single emission line, 
our redshift determination is reliable for the following reasons: 
(1) the absence of other prominent emission lines 
that would be expected if the identified line were not [\ion{O}{2}] 
(such as H$\beta$ or [\ion{O}{3}]), indicating a lower redshift, and 
(2) a close correspondence between the photometric redshift obtained from 
multi-band SEDs and the measured redshift, within the uncertainty of the photometric redshift.

We assign the quality flag `a' to galaxies with secure redshift measurements 
based on clear emission line(s) or absorption feature(s). 
For galaxies where emission or absorption line(s) are faint, 
we assign the quality flag `b'. 
Out of 177 galaxies for which slits were assigned, redshifts could 
be measured for 105 galaxies. 
Among these, 76 were assigned the flag `a', and 29 were assigned the flag `b'.
Therefore, the redshift measurement rate is 66\% 
(44\% for galaxies with the flag `a' only). 
The normalized median absolute deviation (NMAD) of $\Delta z / (1+z)$ for 
galaxies with flag `a' galaxies is 0.024, and NMAD for galaxies with flag `a' or `b' is 
0.030. 
For subsequent analysis, only spectroscopic redshifts with the `a' flag are utilized.

Figure~\ref{fig:szdz} shows the photometric redshift error as a function of 
spectroscopic redshift for spectroscopically confirmed 105 galaxies. 
This figure shows that the photometric redshifts are very accurate 
in the redshift range of $0.8 < z < 1.0$.

The top panel of Figure~\ref{fig:specrat} shows the Subaru $R$-band magnitude 
distribution of our 177  spectroscopic target galaxies (grey histogram) alongside the 
distribution of 76 galaxies whose spectroscopic redshifts were measured with the 
`a' flag (black histogram). 
Given that the $R$-band magnitude was used in determining the priority 
of spectroscopic targets, the number of target galaxies rapidly decrease at $R > 24$ mag. 
While faint galaxies down to $R \sim 27$ mag are included in our target, 
reliable spectroscopic redshift measurements can only be obtained down to $R \sim 24.5$ mag. 

The bottom panel of Figure~\ref{fig:specrat} presents the $R$-band 
magnitude-dependent redshift measurement rate. 
The black symbols and line depict the redshift measurement rate of all galaxies 
with spectroscopic redshifts measured with the flag `a'. 
As can be seen in this figure, the redshift measurement rate declines 
with increasing $R$  magnitude, although it remains at $\sim 50 \%$ down to 
$R \sim 23.5$.
The green and red lines represent the redshift measurement 
rates of blue ($R-z' < 1.5$) and red ($R-z' > 1.5$) galaxies, respectively. 
Although faint red galaxies are included in our target selection, the measurement 
rate sharply declines at $R > 23$ due to the requirement of spectroscopic 
confirmation involving the detection and measurement of continuum and absorption spectral features.

We provide a complete list of galaxies with measured spectroscopic redshifts in 
Table \ref{tab:specztab}, and present several example spectra of galaxies 
in Figure~\ref{fig:spexam}.


\startlongtable
\begin{deluxetable*}{ccccccccc}
\tablenum{1}
\tablecaption{Spectroscopic redshift information from Magellan/IMACS observation\label{tab:specztab}}
\tablewidth{0pt}
\tablehead{
\colhead{ID} & \colhead{R.A. (J2000)} & \colhead{Dec. (J2000)} & 
\colhead{$z_{spec}$} & \colhead{flag} & \colhead{feature} &\colhead{$z_{phot}$} & 
\colhead{$R_{AB}$} & \colhead{Cluster ID} \\
 \colhead{(1)} & \colhead{(2)} & \colhead{(3)} & \colhead{(4)} & 
\colhead{(5)} & \colhead{(6)} & \colhead{(7)} & \colhead{(8)} & \colhead{(9)}
}
\startdata
  244417 &34.05622 & -4.99749 &0.9522 & b & 1 & 0.987 & 23.239 & -- \\
  245714 &34.04414 & -4.99579 &0.8979 & b & 4 &  0.973 &  22.343 & -- \\
  250613 &34.08777 & -4.98057 &0.9432 & b & 1 & 0.900 &  22.865 & -- \\
  251538 &34.08322 & -4.97781 &1.0196 & b & 1 & 0.877 & 22.847 & -- \\
  252172 &34.05969 & -4.97543 &1.0700 & a & 1 &  0.936 & 22.621& -- \\
  255008 &34.04831 & -4.96708 &0.9257 & a & 1 & 0.907  &22.779& -- \\
  256349 &34.04827 & -4.96348 &1.0389  & a & 1 & 0.913 & 22.859& -- \\
  256932 &34.05170 & -4.96175 &0.8841 & a & 1, 2 & 0.895 & 23.411& -- \\
  257855 &34.06900 & -4.95946 &1.0026  & b & 3 & 0.925  &24.687& -- \\
  258156 &34.06051 & -4.95882 & 0.9269 & a & 1 & 0.930   &23.721& -- \\
  260443 &34.05391 & -4.95191 &0.9528  & a & 1 & 0.932  &23.477& OD5 \\
  264268 &34.09649 & -4.94119 &1.0481 & b & 1 & 0.975 & 23.503 & -- \\
  265309 &34.04951 & -4.93914 &0.9568 & b & 1 & 0.910 & 22.562 & OD5 \\
  265593 &34.03121 & -4.93778 &1.0374   & a & 1 & 0.952  &22.870 & -- \\
  267056 &34.12305 & -4.93415& 1.0139  & a & 1 & 0.922 & 22.577& -- \\
  267126 &34.12066 & -4.93504 &1.0142 & b & 1, 3 & 0.912 & 22.747& -- \\
  267460 &34.12469 & -4.93288 &1.0149  & a & 1 & 0.908  &23.876& -- \\
  270520 &34.02529 & -4.92377 &1.0536 & b & 1 & 0.961 & 23.477 & -- \\
  271494 &34.01935 & -4.92221 &1.0542 & b & 1 & 0.968 & 23.819 & -- \\
  271834 &34.04345 & -4.91991 &0.9574 & a & 1 & 0.895 & 22.958& OD5 \\
  273315 &34.10436 & -4.91658 &0.9288 & a & 4 & 0.949  &23.160 & -- \\
  273390& 34.07657 & -4.91531 &0.9571 & a & 1 & 0.914 & 23.570 & OD5 \\
  274769& 34.06047 & -4.91084 &1.0129  & a & 1 & 0.925 & 23.900& -- \\
  275297 &34.10982 & -4.90982 &0.9560 & a & 1 & 0.887  &23.000  & OD5 \\
  275833 &34.06886 & -4.90789 &0.8733 & a & 1, 2 & 0.876  &23.577& -- \\
  276591& 34.11271 & -4.90561 &0.9542 & a & 1 & 0.906 & 24.317& OD5 \\
  277018 &34.12688 & -4.90440 &1.0615 & a & 1 & 0.920  & 24.212& -- \\
  277503 &34.04987 & -4.90333 &1.0612 & b & 1 & 0.959 & 24.173 & -- \\
  278883 &34.07732 & -4.90045 &1.0690  & a & 1 & 0.981  &23.490 & -- \\
  279334& 34.03289 & -4.89895 &0.9539  & a & 1  & 0.885 & 22.620& OD5 \\
  279446 &34.05863 & -4.89776 &1.0624  & a & 1 & 0.941 & 23.287& -- \\
  280515& 34.04927 & -4.89640 & 0.9556 & a & 1, 3 & 0.870 &  22.185& OD5 \\
  280801 &34.05230 & -4.89432 &0.9343 & b & 3 & 0.957 & 22.301& -- \\
  281193 &34.07362 & -4.89516 &1.0132  & a & 1, 3 & 0.916 & 22.428& -- \\
  282124 &34.10719 & -4.88912 &0.9278 & a & 1 & 0.906  &23.826& -- \\
  282407 &34.05711 & -4.89196 & 0.8864 & b & 3 & 0.886 & 22.546 & -- \\
  283044 &34.06182 & -4.88738 &0.9263 & a & 1 & 0.883  &23.017& -- \\
  285119 &34.01437 & -4.88162 &0.9389 & b & 3 & 0.989 & 22.849& -- \\
  285422 &34.02600 & -4.87971& 0.9502 & a & 1 & 0.933 & 23.682& OD5 \\
  285957& 34.10335 & -4.87837 &0.9585 & b & 3 & 0.925  &24.143& OD5 \\
  286455 &34.04077 & -4.87675 &0.9537  & a & 1 & 0.937&  23.174& OD5 \\
  286983 &34.01588 & -4.87473 &0.9497 & a & 1 & 0.931  &23.522& OD5 \\
  287071 &34.04088 & -4.87489 &0.9579 & a & 1, 3  & 0.949 & 24.02 & OD5 \\
  287799& 34.03378 & -4.87290 &0.9480  & a & 1, 3 & 0.914 & 23.686& OD5 \\
  288650 &34.04529 & -4.87064 &0.9539 & b & 1 & 0.910 & 24.021& OD5 \\
  289803& 34.07054 & -4.86697 &1.0437  & a & 1 & 0.933  &23.603& -- \\
  291771 &34.02039 & -4.86163 &0.9328 & b & 3 & 0.903 & 24.088 & -- \\
  293978& 34.04612 & -4.85510 &0.9496  & a & 1 & 0.951  &23.795& OD5 \\
  294085 &34.12790 & -4.85608 &1.0419  & a & 1 & 0.979 & 22.755& -- \\
  295199& 34.08322 & -4.85155 &0.9563  & a & 1 & 0.911  &24.076& OD5 \\
  296511& 34.09816 & -4.84826 &1.0437  & a & 1 & 0.942  &23.701& -- \\
  299918& 34.05652 & -4.84142 &0.9555  & a & 1 & 0.936  &23.490& OD5 \\
  300128& 34.09291 & -4.83823 &0.9501 & a & 1 & 0.923  &23.865& OD5 \\
  300247 &34.05586 & -4.83826 &1.0057 & a & 1 & 0.941  &23.494& -- \\
  302274 &34.09396 & -4.83384 &0.9338 & b & 3 & 0.975 & 23.077& -- \\
  302482& 34.04108 & -4.83204 &0.9541  & a & 1 & 0.912  &23.080& OD5 \\
  302663 &34.07093 & -4.83091 &1.0096 & b & 1 & 0.939 & 24.232 & -- \\
  304508& 34.02532 & -4.82624 &0.8477 & a & 1, 6   & 0.924 & 23.216& -- \\
  304910 &34.09086 & -4.82549 &0.9561 & a & 1 & 0.919  &22.514& OD5 \\
  306231& 34.10122 & -4.82172 &0.8334 & a & 1, 2, 6, 7 & 0.840  & 23.174& -- \\
  306600 &34.10518 & -4.82173 &1.1027 & a & 1 & 0.971  &22.493& -- \\
  307616 &34.13180 & -4.81764 &1.0196  & a & 1 & 0.928  &23.339& -- \\
  307680 &34.12864 & -4.81722 &0.9517 & b & 1 & 0.930 & 24.452 & -- \\
  308869 &34.01946 & -4.81459 &0.8786  & a & 3, 4 & 0.897 & 23.252& -- \\
  309350 &34.01620 & -4.81266 & 0.8940 & b & 1, 3 & 0.906 & 24.062 & -- \\
  310759& 34.07322 & -4.80818  &0.9987  & a & 1, 3 & 0.923 & 24.166& -- \\
  312083 &34.13118 & -4.80484 &0.9069 & b & 1, 5 & 0.911 & 24.309& -- \\
  312773 &34.03048 & -4.80246 &0.9650 & a & 1, 3 & 0.937 & 23.972& -- \\
  314583 &34.12050 & -4.79875 &1.0126  & a & 1 & 0.945  &22.992& -- \\
  314733& 34.13864 & -4.79779 &0.9069  & a & 1, 2, 7 & 0.926  &23.030& -- \\
  315279 &34.02424 & -4.79571 &1.0278 & b & 1 & 0.945 & 24.163& -- \\
  316009& 34.12718 & -4.79397 &0.9187  & a & 1 & 0.939  &24.098& -- \\
  316036 &34.13282 & -4.79524 &0.9087 & a & 1 & 0.910 &22.994& -- \\
  317272 &34.02931 & -4.79094 &1.0240  & a & 1 & 0.909 & 23.649& -- \\
  318139 &34.04814 & -4.78929 &0.9258 & a & 1 & 0.858  &23.853& OD6 \\
  318872 &34.05303 & -4.78640  & 1.0100 & a & 1 &  0.937 & 23.277& -- \\
  320222 &34.13063 & -4.78334  &0.7690 & b & 3, 4 & 0.933 & 22.722  &  -- \\
  321245 &34.09541 & -4.77970 &0.9539 & b & 3 & 0.922 & 23.737 & -- \\
  321954 &34.12612 & -4.77666 &0.8781 & a & 1 & 0.879  &24.020 & -- \\
  322784& 34.08195 & -4.77559 &0.9303 & a & 1 & 0.912  &24.245& OD6 \\
  323555 &34.08519 & -4.77300 &0.9865 & b & 3 & 0.896 & 23.589& -- \\
  326976 &34.09032 & -4.76285 &1.0077  & a & 1, 3 & 0.923  &23.646& -- \\
  327269& 34.11302 & -4.76144 &0.9563  & a & 1, 3 & 0.934  &24.471& -- \\
  329647& 34.02022 & -4.75503 &0.8471  & a & 1, 6 & 0.849  &23.617& -- \\
  329905 &34.09185 & -4.75558 &0.9339 & b & 1, 3 & 0.867 & 24.204 & OD6 \\
  330837& 34.13920 & -4.75191 &0.9560  & a & 1 & 0.912 & 23.437& -- \\
  333013 &34.14636 & -4.74600 &1.0276  & a & 1, 3 & 0.903  &24.321& -- \\
  333403 &34.01790 & -4.74555  &0.9592  & a & 1 & 0.888 & 23.445& -- \\
  333438 &34.07702 & -4.74511 & 0.9835 & a & 1 & 0.935 & 23.087& -- \\
  334775 &34.04690 & -4.74118 &1.0064 & b & 1 & 0.889 & 24.020  & -- \\
  335066 &34.10069 & -4.74156 & 0.9308 & a & 3 & 0.918 & 23.103& OD6 \\
  336397 &34.13436 & -4.73659 &1.0244  & a & 1 & 0.925  &24.004& -- \\
  336469 &34.04198 & -4.73725 &1.0071  & a & 1 & 0.982 & 23.184& -- \\
  337933 &34.01030 & -4.73300 &1.0082  & a & 1 & 0.894 & 22.706& -- \\
  337967 &34.02968 & -4.73159 &0.9578 & a & 1 & 0.927 & 23.930 & -- \\
  338563 &34.02084 & -4.73197 &0.9310 & a & 3, 4 & 0.901  &22.228& OD6 \\
  339103 &34.08369 & -4.72883 &0.9543  & a & 1, 3, 5 & 0.916 & 23.462&  -- \\
  340617 &34.07509 & -4.72401 &0.9271  & a & 1 & 0.909  &24.386& OD6 \\
  343044 &34.07878 & -4.71893 & 0.8430 & b & 3, 4 & 0.926 & 23.281 & -- \\
  343667 &34.09394 & -4.71544 &1.0640  & a & 1 & 0.972 & 24.343& -- \\
  344164 &34.03337 & -4.71439 &0.9320  & a & 3 & 0.883&  23.674& OD6 \\
  345206 &34.10844 & -4.71122 &1.0261  & a & 1 & 0.926 & 23.308& -- \\
  348499 &34.07541 & -4.70168 &0.8301 & a & 1, 6  &0.878  &23.100& -- \\
  350746 &34.06160 & -4.69843 &0.9355 & a & 3, 4 & 0.937 & 22.912& OD6 \\
  350989 &34.05905 & -4.69616 & 0.9411 & b & 3  & 0.930 & 23.312 & -- \\
\enddata
\tablecomments{(1) Object ID;
(2) R. A. (degree);
(3) Declination (degree); 
(4) Spectroscopic redshift;
(5) Spectroscopic flag;
(6) Redshift identification feature -- 1: [\ion{O}{2}], 2: H$\beta$, 3: Ca H$\&$K, 4: 4000 $\rm{\AA}$ break, 5: H$\delta$, 6: [\ion{O}{3}], 7: H$\gamma$;
(7) Photometric redshift;
(8) $R$-band magnitude (AB);
(9) Associated cluster ID}
\end{deluxetable*}

\subsubsection{Slit Loss Correction}
\label{sec:slitloss}

\begin{figure}
    \plotone{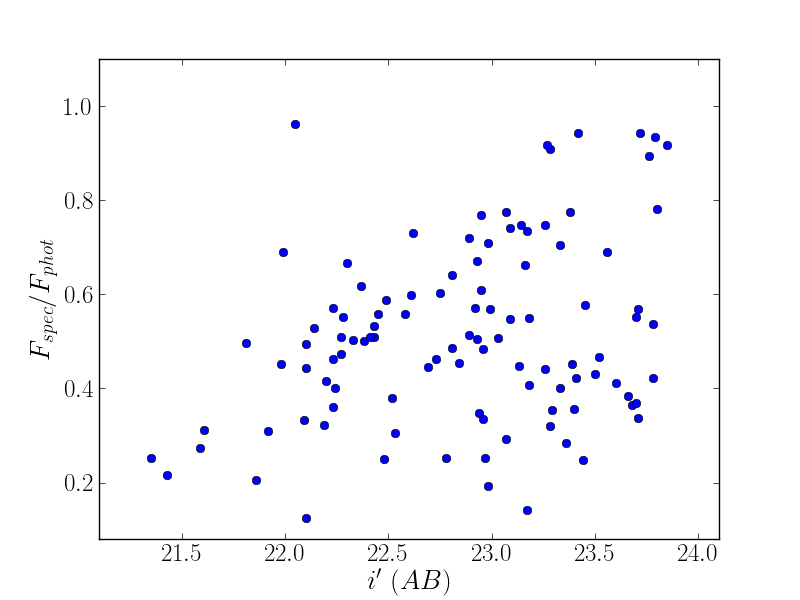}
    \caption{The fraction of fluxes derived from synthetic photometry based 
on the flux-calibrated spectra ($F_{spec}$) compared to the corresponding 
values obtained from Subaru data ($F_{phot}$) in $i'$-band as a funcction of $i'$ magnitude. 
The disparities between these two flux values tend to be more significant 
for brighter galaxies. \label{fig:slitloss}}
\end{figure}

For the spectra of galaxies in Table~\ref{tab:specztab}, 
we compared the synthetic $i'$-band photometry derived from the flux-calibrated 
Magellan spectra with the photometric measurements obtained from Subaru data. 
The differences between these two sets of magnitudes, as shown in Figure~\ref{fig:slitloss}, 
indicate systematic offsets. 
These offsets result from the differences between the width of the spectrograph slit 
and the angular sizes of the observed galaxies. 
This phenomenon, known as slit loss, exhibits a correlation with the magnitude 
of galaxies, with brighter galaxies experiencing more significant slit loss.

\begin{figure}
    \plotone{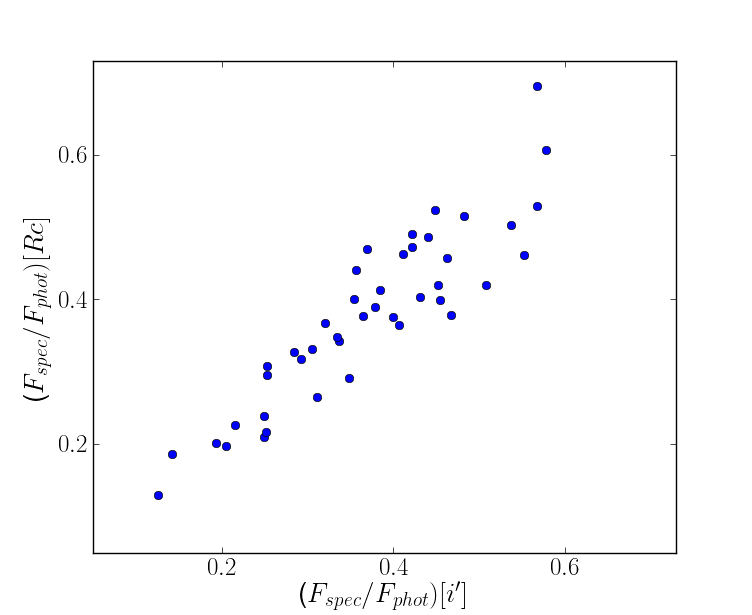}
    \caption{A comparison of the fraction of fluxes ($F_{spec}$/$F_{phot}$) 
obtained in $Rc$-band ($x$-axis) and $i'$-band ($y$-axis). 
These two values are similar, following a one-to-one relation. \label{fig:flratcomp}}
\end{figure}

We extended this comparison to the $Rc$-band photometry and found a similar 
magnitude-dependent trend as well as comparable slit loss values to those 
observed in the $i'$-band. 
The difference between the $Rc$ and $i'$-band correction factors indicates 
the accuracy of flux calibration. 
We found that these two factors agree within 0.05 (Figure~\ref{fig:flratcomp}), 
suggesting flux calibration accuracy of $5 \%$, which becomes to $3 \%$ 
for brighter magnitudes ($i' < 23$).

\begin{deluxetable}{cc}
\tablenum{2}
\tablecaption{The correction factors calculated in the $i'$-band for each galaxy. \label{tab:corfac}}
\tablehead{
\colhead{ID} & \colhead{Correction Factor}}

\startdata
 244417 & 1.67 \\
 245714 & 3.21 \\
 250613 & 2.49 \\
 251538 & 2.41 \\
 252172 & 1.89 \\
 255008 & 3.11 \\
 256349 & 1.79 \\
 256932 & 1.49 \\
 257855 & 1.28 \\
 ..... & ..... \\
\enddata
\tablecomments{Table \ref{tab:corfac} is published in its entirety in the machine-readable format.
      A portion is shown here for guidance regarding its form and content.}
\end{deluxetable}

We provide the correction factors, which correspond to the reciprocal of 
the calculated slit loss values from the $i'$-band analysis, in a separate 
machine-readable table.
A porttion of this table is shown in Table~\ref{tab:corfac}.

\section{Results}
\label{sec:res}

\subsection{Cluster Confirmation}
\label{sec:resspec}

From the spectroscopic redshift data described in Section~\ref{sec:redsh}, 
we confirm two galaxy clusters. 
The first cluster is located at 
R.A.$ = 34.0539$ and Dec.$ = -4.87675$ with a redshift of $z \sim 0.954$. 
The second cluster is located at R.A.$ = 34.0616$ and Dec.$ = -4.73197$ with $z \sim 0.931$. 
We refer to these confirmed clusters as UDS-OD5 and UDS-OD6, respectively.

Initially, among the galaxies with spectroscopic redshifts, we identify potential cluster members 
based on their rest-frame relative velocities, which should fall within $\pm 1500$ km/s and within 
$3 \times R_{200}$ (see Section~\ref{sec:halo} for the derivation of $R_{200}$).

\begin{figure*}
 \plottwo{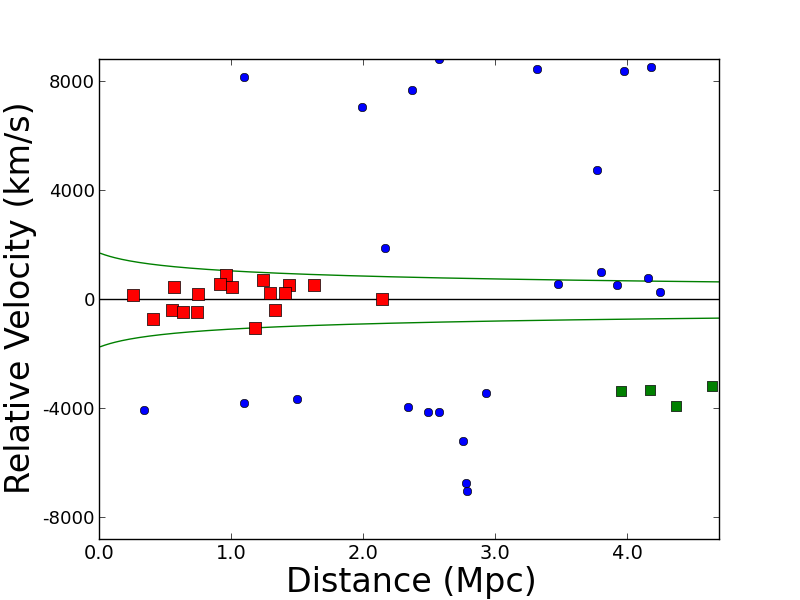}{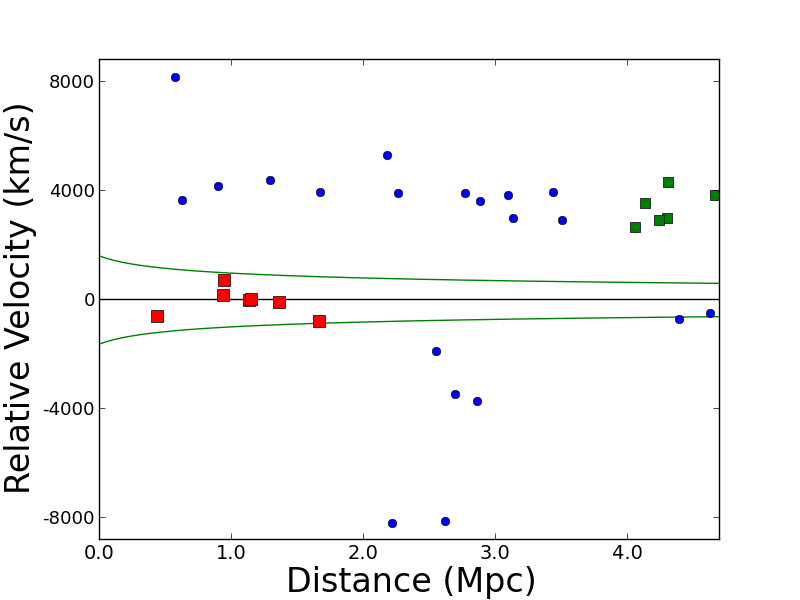}
    \caption{[Left] `Distance--relative velocity' phase diagrams of UDS-OD5. 
Only galaxies with the spectroscopic flag 'a' are shown, and the spectrosopically 
confirmed cluster members are represented by red squares. 
Galaxies represented by green squares are the confirmed members of UDS-OD6.
Galaxies represented by blue dots are not members of either cluster. 
[Right] `Distance--relative velocity' phase diagrams of UDS-OD6, and 
details are same as the left figure.
Some galaxies represented by blue dots at relative velocity of $\sim 4000$ km s$^{-1}$ 
are the galaxies belonging to the large-scale structure connected to UDS-OD5.
\label{fig:phase2}}
\end{figure*}

The final confirmation of cluster members is obtained by investigating the 
`distance--velocity' phase diagram (see Figure~\ref{fig:phase2}). 
We have identified a total of 18 galaxies (with the flag `a') as confirmed members of UDS-OD5, 
and 7 galaxies (with thte flag `a') as confirmed members of UDS-OD6.  
Additionally, there are 3 member galaxies with the flag `b' in UDS-OD5 and 1 member galaxy 
with the flag `b' in UDS-OD6. 
In the subsequent analysis, we will only consider the galaxies with the flag `a'. 


\begin{figure}
  \plottwo{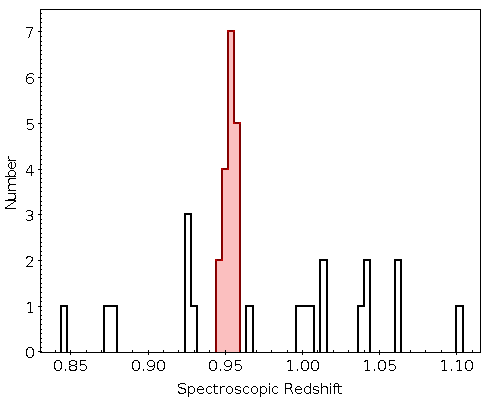}{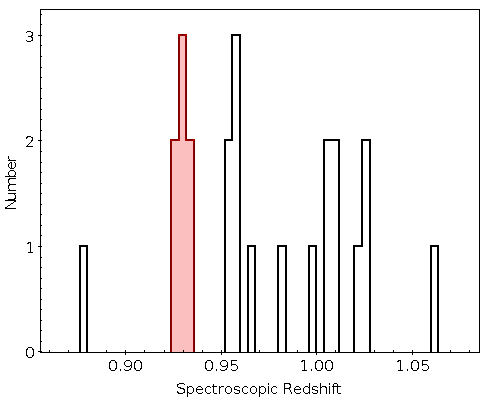}
    \caption{The distributions of spectroscopic redshifts (with the flag `a') 
for galaxies within $3 \times R_{200}$  of UDS-OD5 (left) and of UDS-OD6 (right). 
The red histograms show the redshift distribution of confirmed cluster members. \label{fig:spzdst}}
\end{figure}

In Figure \ref{fig:spzdst}, we show the distributions of spectroscopic redshifts (with the flag `a') 
for galaxies located within $3 \times R_{200}$ from the cluster center. 
In this figure, the red histogram shows the galaxies confirmed as members of each cluster. 
In the redshift distribution of UDS-OD6 (right panel), we can identify a prominent 
peak at $z \sim 0.93$ which shows the galaxies 
belonging to UDS-OD6. 
Additionally, there is a secondary peak at $z \sim 0.95$, as well as two minor peaks at $z > 1$. 
The presence of the second peak at $z \sim 0.95$ in the redshift distribution 
can also be confirmed in the phase diagram in Figure \ref{fig:phase2}. 
In the phase diagram of the right panel in this figure, several galaxies with a relative velocity of 
$\sim 4000$ km s$^{-1}$ belong to the UDS-OD5, and some galaxies with similar relative velocities 
but smaller distances may belong to this second redshift peak. 
Although it is challenging to clearly identify due to the spatial overlap with UDS-OD6, 
some of these galaxies may be part of the filamentary structure connected to UDS-OD5.


\subsection{Halo Properties of Confirmed Clusters}
\label{sec:halo}

From the spectroscopic redshift information of the confirmed member galaxies, 
we can estimate the sizes and masses of the confirmed clusters as follows.  

\subsubsection{UDS-OD5}
\label{sec:od5}

The center of the cluster is determined as the median location of the spectroscopically 
confirmed member galaxies.
This center coincides with the peak of the galaxy number density distribution in this region. 

For the estimation of the halo mass and size, we first calculate the velocity dispersion of the 
cluster from the distribution of spectroscopic redshifts of the confirmed members. 
We employ the gapper estimator \citep{Bee90} for this purpose.
The calculated velocity dispersion is $\sigma_{v}$ = 553 $\pm$ 87 km s$^{-1}$, 
where the error is determined using the $jackknife$ method.
From this $\sigma_{v}$, we estimate $R_{200}$ and $M_{200}$ following the method 
by \citet{dem10}, as follows: 

\begin{equation}
    R_{200}=\frac{\sqrt{3} \sigma_{v}}{10 H(z)}
\label{eq:r200}
\end{equation}

and 
  
\begin{equation}
    M_{200}=3 \frac{\sigma_{v}^{2} R_{200}}{G}.
\label{eq:m200}
\end{equation}

Here, $H(z)$ is the Hubble constant at redshift $z$.
The radius and halo mass of UDS-OD5 are calculated as 
$R_{200} = 799 \pm 126$ kpc and 
$M_{200} = 1.70^{+0.94}_{-0.68} \times 10^{14}$ M$_{\odot}$. 
We also estimate the halo mass from the total stellar mass of galaxies, following \citet{lee19}, and 
applying the weight to cluster candidates as explained in Section~\ref{sec:anaqffof}. 
This method gives the halo mass to be $1.16^{+0.90}_{-0.51} \times 10^{14}$ M$_{\odot}$, 
in agreement with the halo mass estimated from velocity dispersion.

\subsubsection{UDS-OD6}
\label{sec:od6}

The other cluster, UDS-OD6, is located north of UDS-OD5, and at a slightly lower redshift of 
$z \sim 0.931$. 

The velocity dispersion, measured using the Gapper estimator, of the UDS-OD6 is 
$\sigma_{v}$ = 518 $\pm$ 178 km s$^{-1}$. 
The calculated radius and halo mass from the velocity dispersion are $R_{200} = 758 \pm 261$ kpc and 
$M_{200} = 1.42^{+2.03}_{-1.02} \times 10^{14}$ M$_{\odot}$. 
The halo mass from total stellar mass is $1.27^{+0.99}_{-0.56} \times 10^{14}$ M$_{\odot}$. 
From this, we can see that these two clusters, UDS-OD5 and UDS-OD6, have very similar mass and size. 


\subsection{Properties of Candidate Member Galaxies of Confirmed Clusters}
\label{sec:galprop}

\subsubsection{Red Galaxies}
\label{sec:redsq}

\begin{figure}
    \plotone{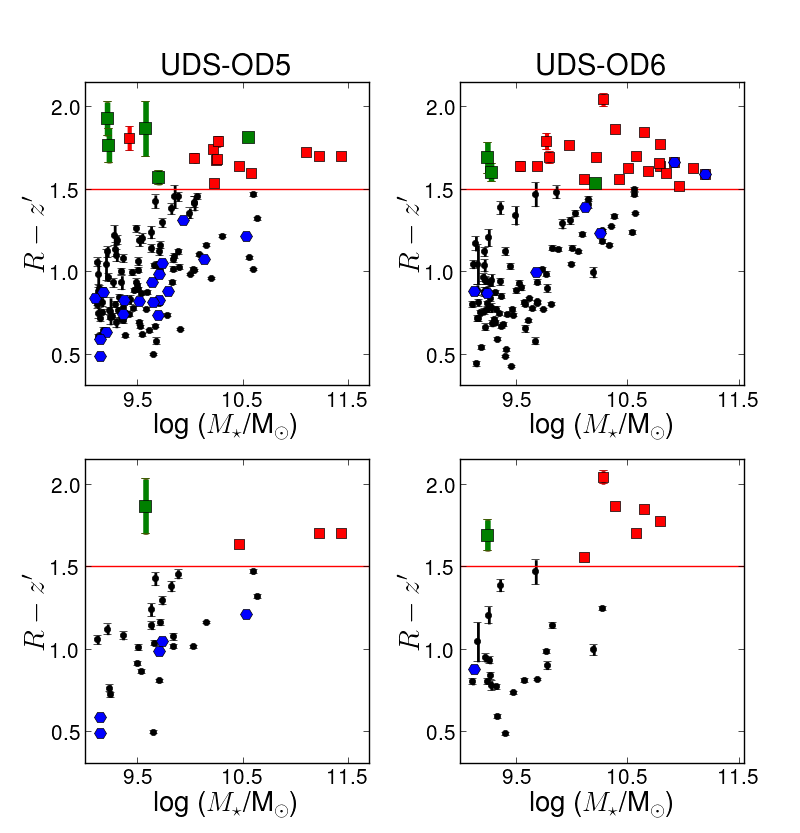}
    \caption{{\bf[Top  panels]:} Diagrams of $R$--$z'$ colors and stellar masses of galaxies within $3 \times R_{200}$ for UDS-OD5 (left) and 
UDS-OD6 (right). {\bf[Bottom  panels]:} Same diagrams for galaxies within $R_{200}$. 
Galaxies whose photometric (or spectroscopic if available) redshifts are within the range of the cluster are shown. 
Blue hexagons are spectroscopically confirmed galaxies and black small circles are galaxies with photometric redshift only. 
Red squares are red quiescent galaxies with $R$--$z'$ color greater than 1.5 and green squares are dusty star-forming galaxies 
with red color ($R$--$z' > 1.5$).  \label{fig:redsq}}
\end{figure}

Figure \ref{fig:redsq} shows the correlation between stellar mass and $R$--$z'$ color of galaxies whose 
redshifts are within the photometric redshift error range ($\Delta z /(1+z_{cl}) = 0.028$). 
The top panels of this figure show galaxies located within $3 \times R_{200}$ from the cluster center, and 
the bottom panels show galaxies within $R_{200}$.  
In this figure, red squares are red quiescent galaxies whose $R$--$z'$ color 
is redder than 1.5. This color cut is shown as a red horizontal line in each panel. 
There is a well-developed red sequence in the regions containing both clusters, especially 
for massive galaxies (M$_{*} > 10^{10}$ M$_{\odot}$).
Not all of the red galaxies are quiescent. From the SED-fitting results, we find some of these red 
galaxies are dusty SF galaxies (shown as green squares in this figure). 
In this figure, blue hexagons are spectroscopically confirmed galaxies. 
As can be seen in the upper right panel of Figure~\ref{fig:redsq}, two red quiescent galaxies are 
spectroscopically confirmed as cluster members of the UDS-OD6 while the association of the other 
red sequence galaxies to the clusters is determined from photometric redshifts, the proximity to 
the overdense areas, and a statistical inference.

\begin{rotatetable}
\begin{deluxetable*}{ccccccccc}
\tabletypesize{\scriptsize}
\tablenum{3}
\tablecaption{Properties of confirmed clusters\label{tab:cluster}}
\tablehead{
\colhead{Name} & \colhead{Redshift} & \colhead{$\sigma_{gap}$\tablenotemark{a}} & \colhead{$R_{200}$} & 
\colhead{$M_{200,v}$\tablenotemark{b}} & \colhead{$M_{200,s}$\tablenotemark{c}} & \colhead{QF$_{1}$\tablenotemark{d}} & 
\colhead{QF$_{2}$\tablenotemark{e}} & \colhead{FoF} \\
\colhead{} & \colhead{} & \colhead{(km/s)} & \colhead{(kpc)} & \colhead{(M$_{\odot}$)} & 
\colhead{(M$_{\odot}$)} & \colhead{} & \colhead{} & \colhead{} 
}
\colnumbers
\startdata
 UDS-OD5 & 0.954 $\pm$ 0.003 & 553 $\pm$ 87 & 799 $\pm$ 126 & 1.70$^{+0.94}_{-0.68}$e+14 & 1.16$^{+0.90}_{-0.51}$e+14 & $0.18_{-0.054}^{+0.11}$ & $0.094_{-0.032}^{+0.11}$ & $0.056^{+0.013}_{-0.028}$ \\
 UDS-OD6 & 0.931 $\pm$ 0.003 & 518 $\pm$ 178 & 758 $\pm$ 261 & 1.42$^{+2.03}_{-1.02}$e+14 & 1.27$^{+0.99}_{-0.56}$e+14 & $0.47_{-0.12}^{+0.13}$ & $0.38_{-0.11}^{+0.14}$ & $0.038^{+0.019}_{-0.0079}$ \\
\enddata
\tablecomments{(1) Cluster name;
(2) Cluster redshift;
(3) Velocity dispersion; 
(4) Cluster radius ($R_{200}$);
(5) Cluster mass within $r \leq R_{200}$ derived from velocity dispersion;
(6) Halo mass derived from total stellar mass using \cite{lee19} relation;
(7) Quiescent Fraction calculated using method 1;
(8) Quiescent Fraction calculated using method 2;
(9) FoF fraction}
\end{deluxetable*}
\tablenotetext{a}{Derived using gapper estimator \citep{Bee90}}
\tablenotetext{b}{Halo mass derived from $\sigma_{gap}$ using \cite{dem10} relation}
\tablenotetext{c}{Halo mass derived from total stellar mass using \cite{lee19} relation}
\tablenotetext{d}{Quiescent Fraction calculated using method 1}
\tablenotetext{e}{Quiescent Fraction calculated using method 2}
\end{rotatetable}

\subsubsection{Spatial distribution of galaxies of various types}
\label{sec:anavar}

\begin{figure}
    \plotone{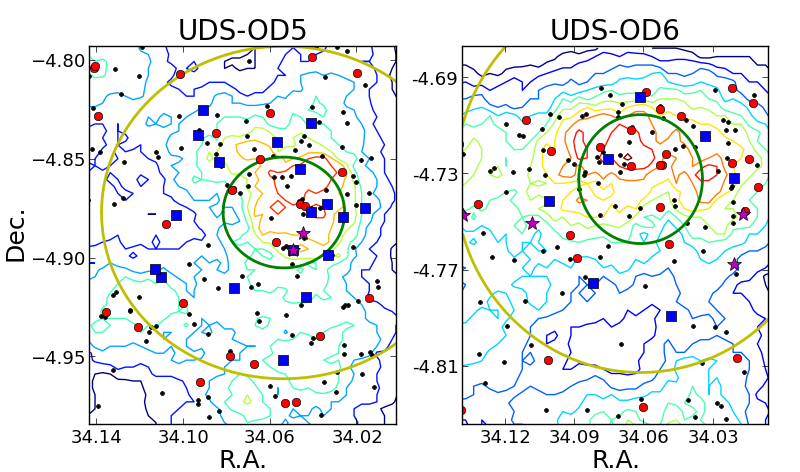}
    \caption{The location of various types of galaxies, plotted over the density maps of UDS-OD5 (left) 
and UDS-OD6 (right). 
Red circles show the locations of red (sequence) galaxies. 
Black dots show the location of galaxies whose photometric redshifts are within the photometric redshift 
error range ($\Delta z /(1+z_{cl}) = 0.028$) of each cluster and blue squares show spectroscopically confirmed galaxies. 
Purple stars show the MIPS-detected galaxies. 
The green and yellow circles in each panel show the $R_{200}$ and $3 \times R_{200}$ of each cluster.  \label{fig:denm}}
\end{figure}

Figure~\ref{fig:denm} shows the spatial distribution of various types of 
galaxies in and around UDS-OD5 (left panel) and UDS-OD6 (right panel), 
plotted over density maps. 
Here, the density map is drawn using galaxies whose (spectroscopic or photometric) 
redshifts are within the photometric redshift error range ($\Delta z /(1+z_{cl}) = 0.028$). 

In this figure, red circles show the locations of red (sequence) galaxies. 
We can see that there is a concentration of red (sequence) galaxies near the 
cluster center in the case of UDS-OD6. 
Black symbols are galaxies within the photometric redshift error range, and 
blue squares are spectroscopically confirmed galaxies among these. 
Purple stars are MIPS-detected galaxies which are dusty star-forming galaxies. 
Interestingly, near the cluster center of UDS-OD5 (left panel), there are concentration of various 
types of galaxies --- red (sequence), SF, and dusty SF galaxies. while red (sequence) 
galaxies are more dominant in the central region of UDS-OD6 (right panel). 

\section{Analysis}
\label{sec:anal}

\subsection{Correlation between the Quiescent Galaxy Fraction and the Large Scale Environment}
\label{sec:anaqffof}

In our previous work, we found that there is a correlation between the QF 
of high-redshift ($0.65 \le z \le 1.3$) galaxy clusters and the extent of 
large scale structure surrounding the clusters \citep{lee19}.
Now, we investigate the QFs of the two newly confirmed clusters and 
large-scale overdensity surrounding them. 
For this analysis, we use both spectroscopic and photometric redshifts.

A galaxy is classified as quiescent if its specific SFR is smaller than $1/(3 \times t(z))$, where 
$t(z)$ is the age of the Universe at redshift $z$  \citep{dam09,lee15}. 
Then, the quiescent galaxy fraction of a cluster is defined as 
the fraction of quiescent galaxies among total cluster member galaxies within $R_{200}$. 
In this work, we applied a stellar mass cut of log ($M_{*}$/M$_{\odot}) \geq 9.1$, 
which is the stellar mass completeness limit, calculated in \citet{lee15}. 

We calculate the QF for each cluster in two ways. 
First, it is calculated by applying the weight ($w$) values which are derived as explained 
below \citep[in a similar way as in][]{lee19} to all galaxies within photometric redshift 
range ($\Delta z \leq 0.028 \times (1+z)$). 
The weight, $w$ is computed  as the number of spectroscopically confirmed cluster 
member galaxies divided by the sum of the number of spectroscopic members and 
outliers. 
For spectroscopically confirmed cluster members, we assign $w=1$.
The QF values calculated in this way (method 1) are 
$0.18_{-0.054}^{+0.11}$ for UDS-OD5 and $0.47_{-0.12}^{+0.13}$  for UDS-OD6. 

Here, the errors are the $68.3 \%$ confidence levels, which have been determined 
employing the method outlined in \citet{cam11}.

\begin{figure}
 \plottwo{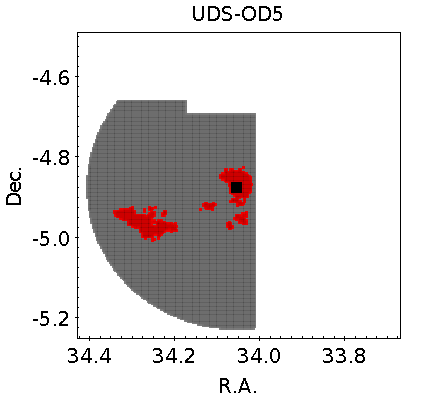}{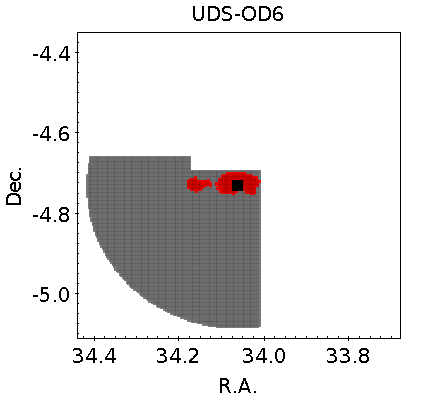}
    \caption{Distribution of the overdense ($\ge 2\sigma$) large-scale structure (shown as red regions) near 
UDS-OD5 (left) and UDS-OD6(right). The grey region shows the total area whose boundary is set by either 
10 Mpc radius from the cluster center or by the survey boundary. The black square in each panel 
shows the location of the cluster center. \label{fig:fofmap}}
\end{figure}

The second way (method 2) we used is to apply the background correction to the 
quiescent galaxy numbers of clusters by subtracting the quiescent galaxy fraction of 
field galaxies in the same redshift range to calculate the corrected quiescent 
galaxy number of each cluster, $N_{qs,cl,excess}$, as follows:

\begin{equation}
    N_{qs,cl,excess}= N_{qs,cl,raw} - n_{qs,fld} \times A_{cl},
\label{eq:clqscal}
\end{equation}

where $N_{qs,cl,raw}$ is the number of quiescent galaxies within $R_{200}$ of 
the cluster, $n_{qs,fld}$ is the surface number density of quiescent galaxies in the field region, 
and $A_{cl}$ is the surface area of the cluster. 
The quiescent galaxy fraction values calculated in this way (method 2) are 
$0.094_{-0.032}^{+0.11}$ and $0.38_{-0.11}^{+0.14}$ for UDS-OD5 and 
UDS-OD6, respectively. 

In both methods, we use both the spectroscopically confirmed memebers and 
the candidate member galaxies with photometric redshift only.
While we found and confirmed the cluster members out to $3 \times R_{200}$, we calculate 
the QF within $R_{200}$ for consistent comparison with our previous work \citep{lee15}. 
The QF values tend to decrease when we use a larger radius, and this tendency was more 
significant for UDS-OD6.
Both methods give higher QF for UDS-OD6. 

As shown in the right panel of Figure~\ref{fig:spzdst}, the spectroscopic redshift distribution of 
UDS-OD6 has several minor peaks in addition to the peak at the redshift of UDS-OD6. 
This might raise a question about the possible contamination of background galaxies belonging to 
minor peaks in calculating QF. 
However, our QF calculation methods can minimize this effect by applying a weight based on 
the spectroscopic redshifts (method 1), or by subtracting QF of field galaxies (method 2) 
minimizing the contamination effect in measuring QF. 
Also, the effect of possible remaining contamination from background galaxies, if any, would 
have made the QF underestimated (because the QF of field galaxies is lower than the QF of clusters), 
which strengthens our conclusion about the correlation between QF and FoF fraction.

To quantify the extent of large-scale structure surrounding the clusters, 
we use the friends-fo-friends fraction (FoF fraction) following \citet{lee19}. 
The FoF fraction is defined as the fraction of (2-dimensional) overdense regions which 
are connected to a cluster over the total area within 10 Mpc radius around the cluster.  
Here, the overdense region is where the galaxy number density is greater than $2 \sigma$ 
from the mean background number density.
And, whether this overdense region is connected to the cluster is determined using FoF algorithm.
So, if there are more overdense structures connected to the cluster, it has a larger FoF fraction. 
To calculated the FoF fraction, we use both photometric and spectroscopic redshifts.
The results of FoF fraction measurement are shown in Figure~\ref{fig:fofmap}. 
In this figure, the red-colored region shows the area where the connected overdense LSS and the black square 
symbol shows the location of each cluster center.

\begin{figure}
    \plotone{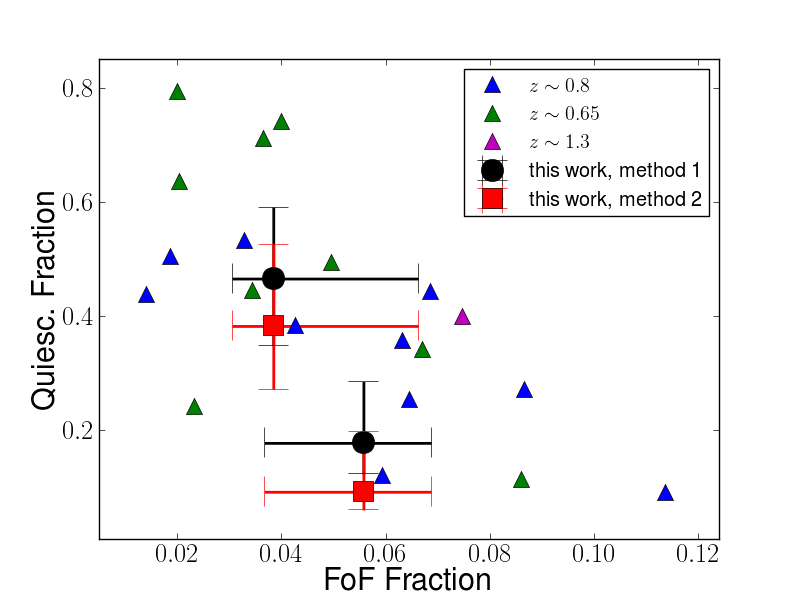}
    \caption{The correlation between quiescent galaxy fractions and FoF fractions of high redshift 
\edit1{($0.65 \le z \le 1.3$)} galaxy clusters. 
Green, blue and purple triangles show the values of galaxy clusters at $z \sim 0.65$, 0.8, and 1.3, respectively from 
\citet{lee19}. Black circles and red squares show the values of this study. 
The black circles and red squares show the QF values calculated following the methods 1 and 2, respectively 
(Please refer to the corresponding text for the explanation of each calculating method). 
Two clusters newly found in this study are consistent with the anti-correlation found in our previous work \citep{lee19}. \label{fig:qffof}}
\end{figure}

The FoF fractions of UDS-OD5 and UDS-OD6 are $0.056^{+0.013}_{-0.028}$ and 
$0.038^{+0.019}_{-0.0079}$, respectively. 
As can be seen in Figure~\ref{fig:fofmap}, the FoF fractions cannot be measured within a 
full circle of 10 Mpc radius for both clusters, because these clusters are located near survey 
boundary.
The uncertainty of FoF fraction is measured through Monte Carlo simulation. 
We assume the FoF fraction value in $unseen$ region follows similar distribution as 
the FoF values found in our previous work \citep[][please refer Figure~\ref{fig:qffof}]{lee19}. 
Then we randomly pick FoF value of $unseen$ region, assuming gaussian distribution and 
calculte the $total$ FoF fraction combining this randomly picked value and the measured FoF fraction. 
We repeat this procedure 1000 times, and the uncertainty is determined as the $68 \%$ confidence 
interval.

These newly-found two clusters have different QFs, while being at similar redshifts and having similar masses.
Their large scale environments as judged from the FoF fractions may be different but the two FoF values are 
within 1-$\sigma$ errors. 
In Figure~\ref{fig:qffof}, we show the correlation between QFs and FoF fractions 
of these two clusters (black circles and red squares) along with the clusters from our previous work \citep{lee19}. 
In this figure, we can see that these newly-found two clusters do not behave differently from the anti-correlation 
between QF and FoF fraction found in \citet{lee19}. 

\subsection{Cluster Stellar Mass Function}
\label{sec:anaassem}

\begin{figure}
    \plotone{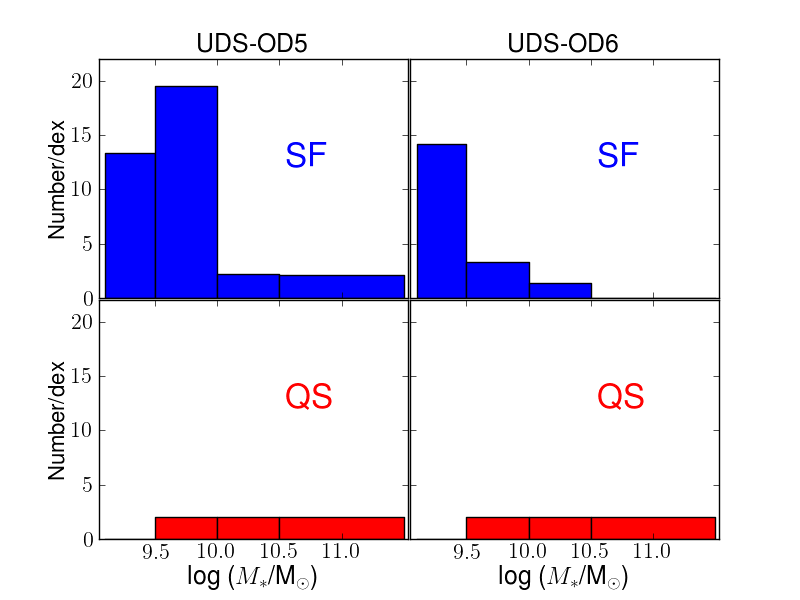}
    \caption{The stellar mass functions of star-forming (top panels) and quiescent (bottom panels) galaxies 
of UDS-OD5 (left panels) and UDS-OD6 (right panels). Candidate cluster members without spectroscopic confirmation are 
included with appropriate weights applied. \label{fig:msftn}}
\end{figure}

Now, we investigate the stellar mass function of confirmed clusters. 
Figure~\ref{fig:msftn} shows the stellar mass functions of quiescent (red histograms) and star-forming galaxies 
(blue histograms) which are within $R_{200}$ of UDS-OD5 (left panels) and of UDS-OD6 (right panels). 
In calculating these stellar mass functions of cluster galaxies, we apply appropriate weights 
(see Section~\ref{sec:anaqffof}) to galaxies with only photometric redshifts.

As can be seen in this figure, the stellar mass functions of quiescent galaxies in UDS-OD5 and UDS-OD6 
are very similar (bottom panels). 
However, the stellar mass functions of star-forming galaxies of these two clusters are largely different (top panels). 
The stellar mass function of star-forming galaxies of UDS-OD5 shows a clear excess compared 
to that of UDS-OD6.
This excess of star-forming galaxies of UDS-OD5 leads to a lower QF than UDS-OD6. 
Considering the fact that there are more overdense structures surrounding UDS-OD5 than UDS-OD6, 
we can speculate that these additional star-forming galaxies have accreted from the surrounding 
overdense structure, in good accordance with our suggested WFM. 

Galaxies become quiescent through two distinct mechanisms: in-situ and accretion. 
The in-situ process suggests that the evolution of quiescent galaxies is dependent on their mass, 
with more massive galaxies transitioning to the red sequence at higher redshifts \citep{san09}. 
The formation of quiescent galaxies, whether through in-situ processes or accretion, exhibits distinct 
dependencies on redshift and environment.
In the early universe, in-situ star formation prevails, causing galaxies to transition from active 
star formation to quiescence over time. 
Massive galaxies become quenched and red as early as $z>2$ \citep{kri08,gab12}.
On the other hand, accretion becomes more significant in the late-time universe, with a potential increase 
in its contribution over cosmic time, especially for low-mass galaxies \citep{pen15,lee15,cer16}. 
Dense environments and large-scale structures, such as galaxy clusters, play crucial roles in facilitating accretion processes.

In Figure~\ref{fig:msftn}, it is evident there is no difference in the stellar-mass functions of quiescent galaxies 
between the two clusters. 
Additionally, it is noteworthy that there are no quiescent galaxies with log $M_{*}$/M$_{\odot} < 9.5)$ in either cluster, 
suggesting that the environmental factors may not be fully at plat yet. 
In case of the stellar mass functions of SF galaxies, we observe a higher presence of low-mass 
SF galaxies in OD5. 
We hypothesize that these galaxies, relatively recently accreted from the surrounding large-scale structure, 
will cease star formation over time due to environmental factors and eventually join the red sequence. 
Consequently, over time, OD5 is expected to harbor a larger number of red sequence galaxies compared to OD6.

\begin{figure}
    \plotone{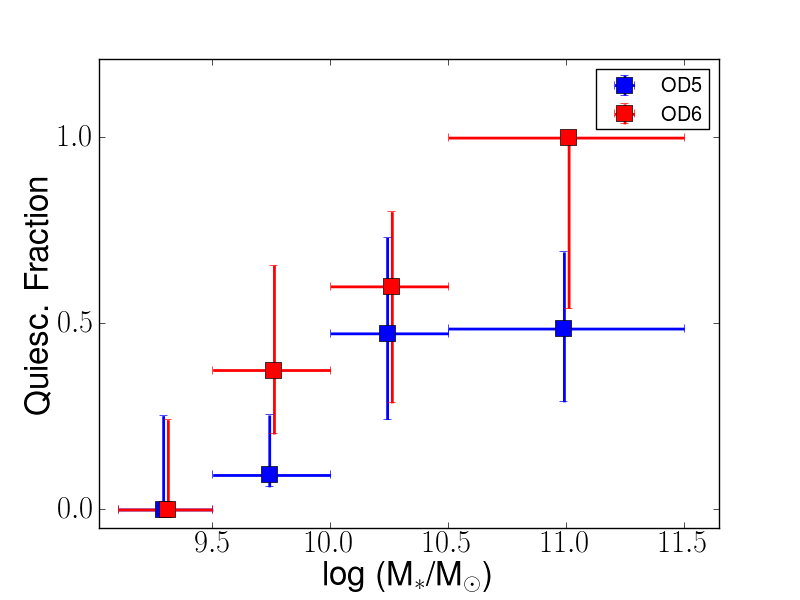}
    \caption{The stellar-mass dependent quiescent galaxy fraction within $R_{200}$ of UDS-OD5 (blue squares) 
and of UDS-OD6 (red squares). Candidate cluster members without spectroscopic confirmation are included 
with appropriate weights applied. As can be seen here, OD6 has larger QFs in three stellar mass bins. \label{fig:qfms}}
\end{figure}

Figure~\ref{fig:qfms} shows the stellar-mass dependent QFs of UDS-OD5 (blue histograms) and UDS-OD6 
(red histograms), 
within $R_{200}$ of each cluster. 
In both clusters, we can see a clear stellar-mass dependence of cluster QF, in a 
sense that QF is higher for more massive galaxies. 
Also, we can see that quiescent galaxy fractions of UDS-OD5 are lower than UDS-OD6 in three (out of four) 
mass bins, which leads to the lower total quiescent galaxy fraction of UDS-OD5. 
As mentioned above, this difference is speculated to be originated by recently fallen galaxies from 
surrounding environment.

Stellar-mass dependent trend in QFs shown in Figure~\ref{fig:qfms} indicates that QF would increase 
if we apply a higher stellar-mass cut for both UDS-OD5 and UDS-OD6.  
But, the change of stellar-mass cut would not change our main conclusion, 
since this stellar-mass dependent trend is shown in both clusters in a similar manner.


\section{Conclusion}
\label{sec:conc}

We conducted spectroscopic observations of galaxies with a redshift of $z \sim 0.95$ 
in and around galaxy cluster candidates identified in the UDS field using Magellan/IMACS. 
We present the flux-calibrated spectra and redshifts of 105 galaxies for which redshift 
measurements were possible. 
Through the analysis of these 105 spectroscopic redshifts,  
we have identified two galaxy clusters: one at $z \sim 0.95$ (UDS-OD5) and 
another at $z\sim0.93$ (UDS-OD6). 
Additionally, by integrating spectroscopic redshift information with multi-band photometric data, 
we have made the following findings:  

\begin{enumerate}

\item [1.] The identified clusters exhibit halo masses of $M_{200} \sim 10^{14}$ M$_{\odot}$, 
and sizes of $R_{200} \sim 0.8$ Mpc. 
Specifically, for UDS-OD5, the estimated halo mass is $M_{200} = 1.7 \times 10^{14}$ M$_{\odot}$ 
with a corresponding size of $R_{200} = 800$ kpc, derived from spectroscopic redshift information. 
Similarly, for UDS-OD6, the values are  $M_{200} = 1.4 \times 10^{14}$ M$_{\odot}$ and $R_{200} = 760$ kpc. 
We have also estimated the halo masses using the correlation between total stellar mass and halo mass 
presented in \citet{lee19}. 
This method yields halo mass estimates of $M_{200} = 1.2 \times 10^{14}$ M$_{\odot}$ 
for UDS-OD5 and $1.3 \times 10^{14}$ M$_{\odot}$ for UDS-OD6, which are consistent 
with the values derived from spectroscopic redshifts.

\item [2.] Despite their similar masses, sizes, and redshifts, these two clusters 
exhibit distinct star formation properties.  
While a well-formed red sequence exists, UDS-OD5 shows a prevalence of blue, star-forming galaxies. 
The quiescent galaxy fraction for UDS-OD5 is estimated to be $0.18 \pm 0.097$ (method 1) or 
$0.094 \pm 0.071$ (method 2). 
In contrast, UDS-OD6 displays a higher quiescent fraction of $0.47 \pm 0.21$ (method 1) or 
$0.38 \pm 0.20$ (method2) with many red sequence galaxies, 
particularly concentrated in the cluster's central region. 

\item [3.] While the limited coverage of the surrounding area introduces some uncertainty, 
it appears that these two clusters inhabit different large-scale environments. 
The FoF fraction values for UDS-OD5 and UDS-OD6 are $0.056^{+0.013}_{-0.028}$ and 
$0.038^{+0.019}_{-0.0079}$, respectively. 
These FoF values and QF values of these two clusters are consistent with 
the (anti-)correlation observed in high-redshift \edit1{($0.65 \le z \le 1.3$)} clusters \
as reported in \citet{lee19}, supporting our suggested WFM.   

\item [4.] We examine the stellar mass function of cluster galaxies for distinct 
populations of quiescent and star-forming galaxies. 
Both UDS-OD5 and UDS-OD6 show similar quiescent galaxy stellar mass functions. 
However, there is a discrepancy in the stellar mass functions of star-forming galaxies 
between the two clusters. 
UDS-OD5 displays an overabundance of star-forming galaxies compared to UDS-OD6. 
We propose that this excess of star-forming galaxies found in UDS-OD5 stems from recent 
infall of galaxies from the surrounding large-scale environment, 
providing further support for our WFM. 

\end{enumerate}
 
Our spectroscopic data will be valuable for future studies of galaxies in the UDS field. 
The properties of the two newly confirmed clusters at $z \sim 0.95$ are consistent 
with our previously suggested WFM. 
Future studies on this correlation over a larger area and employing more extensive multi-object spectroscopy 
would provide a deeper understanding about the interplay between galaxy clusters and LSS, as well as 
its impact on the evolution of galaxy clusters. 

\begin{acknowledgements}

This work was supported by the National Research Foundation of Korea (NRF) grant No. 
2020R1I1A1A01060310, No. 2020R1A2C3011091 and No. 2021M3F7A1084525, funded by the Korea government (MSIT).
This paper includes data gathered with the 6.5 m Magellan Telescopes located at Las Campanas
Observatory, Chile, and from the UKIDSS \citep{law07} UDS \citep{alm07} project. 
UKIDSS uses the UKIRT Wide Field Camera \citep[WFCAM;][]{cas07}. The photometric system is described in 
\citet{hew06}, and the calibration is described in \citet{hod09}. The pipeline processing and 
science archive are described in \citet{ham08}. \\

\end{acknowledgements}

\vspace{5mm}
\facilities{Magellan(IMACS), UKIRT(WFCAM), Subaru(SUPRIMECAM), Spitzer(IRAC)}

\appendix

\restartappendixnumbering

\section{Flux Calibrated Galaxy Spectra}

We reduced the Magellan MOS spectra using the COSMOS2 software. 
This reduction process encompassed several key steps, namely wavelength calibration, 
flat fielding, bias subtraction, and sky subtraction. 
This process resulted in a set of processed 2D spectra, which were then combined to generate 1D spectra.

For these resulting 1D spectra, we conducted flux calibration by referencing a star's spectrum 
specifically included in the Magellan mask for calibration purposes. 
This calibration was established by comparing the Magellan spectrum of the star 
with its counterpart from the Sloan Digital Sky Survey (SDSS).

\begin{figure}
    \includegraphics[width=\columnwidth]{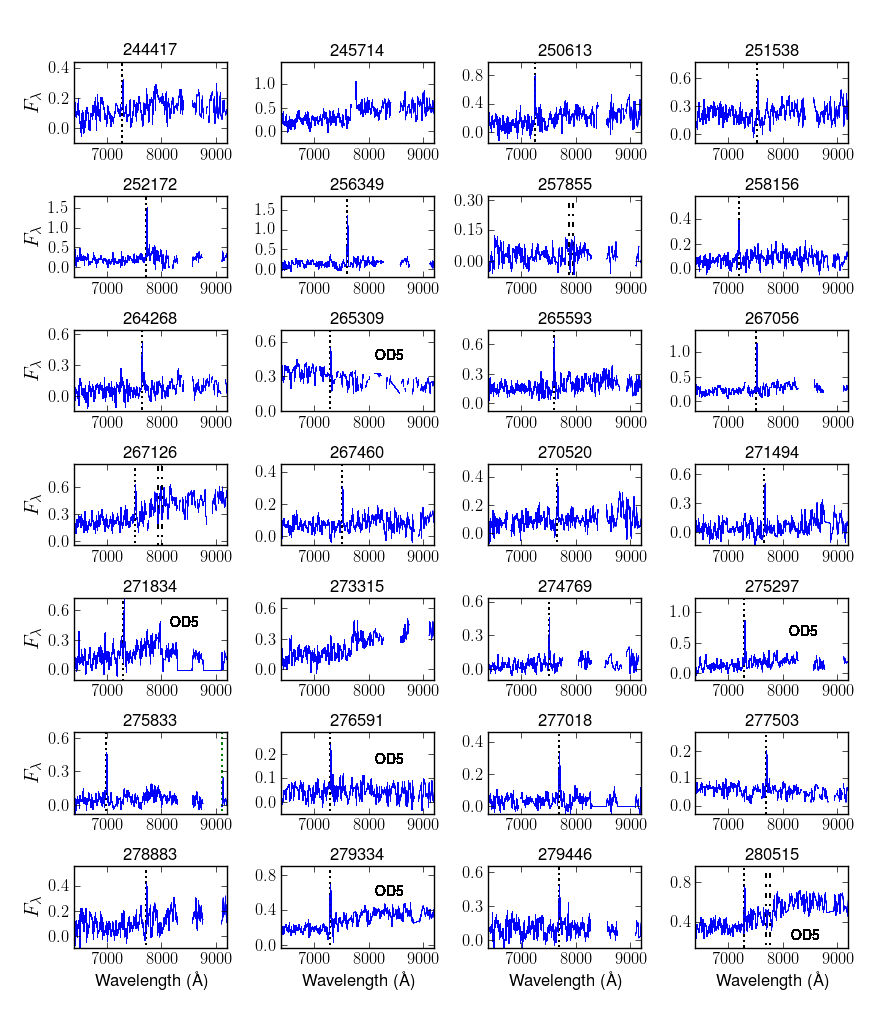}
    \caption{Spectra of spectroscopically confirmed from Table~\ref{tab:specztab}.
The $y$-axis is in the unit of $10^{-17}$ $erg$ $s^{-1}$$cm^{-2}$\AA$^{-1}$. 
The colored dotted lines show the locations of various emission lines as follows: 
black-[\ion{O}{2}], green-H$\beta$, red-H$\delta$, orange-[\ion{O}{3}], magenta-H$\gamma$.
The black dot-dashed lines show the location of redshifted Ca H$\&$K absorption lines. \label{fig:clspecs}}
\end{figure}

We provide the flux calibrated spectra of all the galaxies from Table~\ref{tab:specztab} 
in a machine-readable form (Table~\ref{tab:calflx}).
We also show spectra of galaxies from Table~\ref{tab:specztab} in 
Figure~\ref{fig:clspecs}.

\begin{figure}

    \includegraphics[width=\columnwidth]{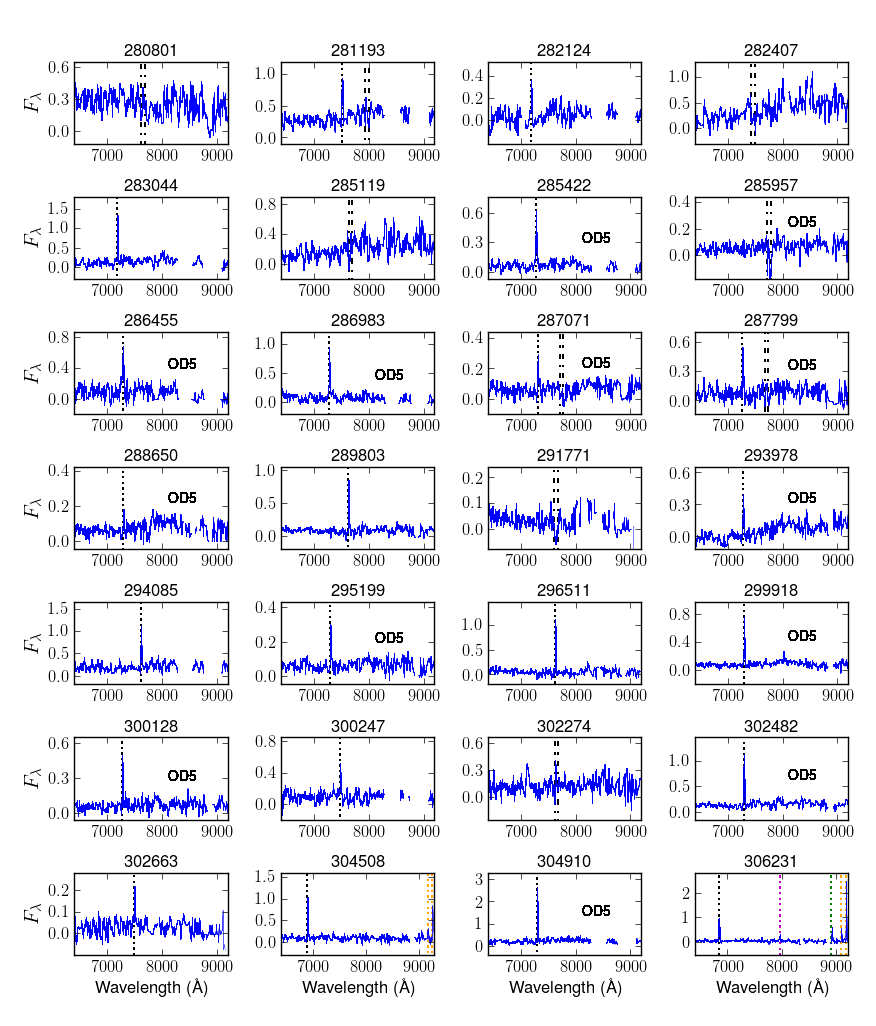}\\

    {\bf Figure A1.} continued\\
\end{figure}

\begin{figure}
    \includegraphics[width=\columnwidth]{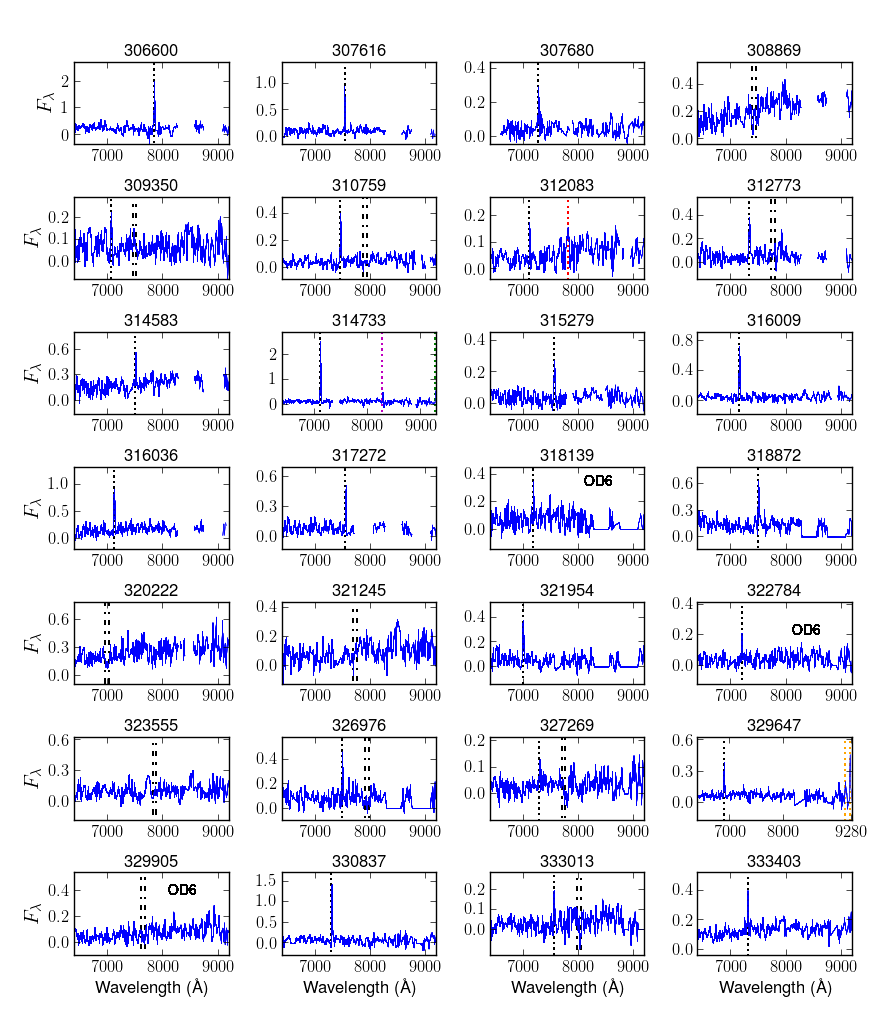}\\
    
    {\bf Figure A1.} continued\\
\end{figure}

\begin{figure}
    \includegraphics[width=\columnwidth]{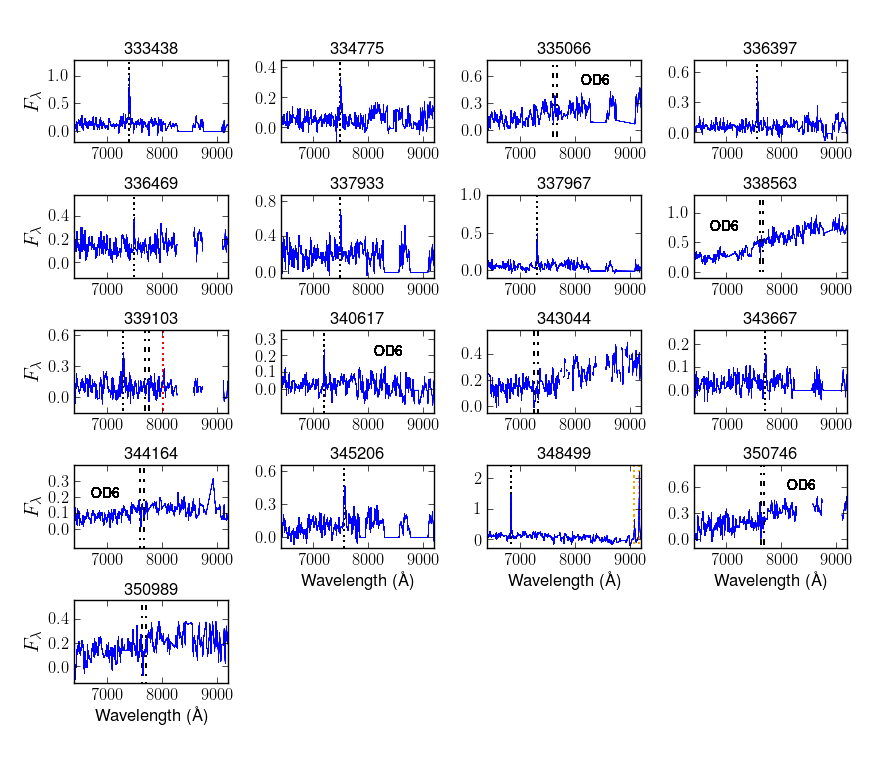}\\
    
    {\bf Figure A1.} continued\\
\end{figure}

\begin{deluxetable*}{ccc}
\tabletypesize{\small}
\tablecaption{Flux-calibrated spectrum of the galaxy with ID number 314733. 
(Abbreviated. Full spectra of all galaxies are provided in FITS format.) \label{tab:calflx}}
\tablehead{
\colhead{Wavelength (\AA)} & \colhead{Flux ($10^{-17}$ $erg$ $s^{-1}$$cm^{-2}$\AA$^{-1}$)} & \colhead{Flux Error} ($10^{-17}$ $erg$ $s^{-1}$$cm^{-2}$\AA$^{-1}$)}

\startdata
5698.0 & 0.0000 & 1.6128E-4 \\
5700.0 & 0.0992 & 1.6016E-4 \\
5702.0 & 0.1067 & 1.6468E-4 \\
5704.0 & 0.0312 & 1.7731E-4 \\
5706.0 & 0.1007 & 1.6318E-4 \\
5708.0 & 0.0940 & 1.5242E-4 \\
5710.0 & 0.2542 & 1.4999E-4 \\
5712.0 & 0.2646 & 1.5910E-4 \\ 
5714.0 & 0.2750 & 1.6044E-4 \\ 
5716.0 & 0.2043 & 1.6650E-4 \\
5718.0 & 0.0931 & 1.7095E-4 \\
5720.0 & 0.0878 & 1.5808E-4 \\
5722.0 & 0.0894 & 1.6296E-4 \\
5724.0 & 0.1374 & 1.7439E-4 \\
5726.0 & 0.1259 & 1.5464E-4 \\
5728.0 & 0.1290 & 1.5850E-4 \\
5730.0 & 0.1334 & 1.5312E-4 \\
 ..... & ..... & ..... \\
\enddata
\tablecomments{Table \ref{tab:calflx} is published in its entirety in the machine-readable format.
      A portion is shown here for guidance regarding its form and content.}
\end{deluxetable*}


\bibliography{uds_draft_slee}{}
\bibliographystyle{aasjournal}


\end{document}